\documentstyle[12pt,aaspp4]{article}

\begin{document}

\title{Main Sequence Masses and Radii from Gravitational Redshifts}

\author{Ted von Hippel\altaffilmark{1}}
\affil{Department of Astronomy, University of Wisconsin, Madison, WI 53706,
USA}
\authoremail{ted@noao.edu}
\altaffiltext{1}{address: WIYN Telescope, NOAO, PO Box 26732, Tucson, AZ
85726-6732, USA}

\begin{abstract}

Modern instrumentation makes it possible to measure the mass to radius
ratio for main sequence stars in open clusters from gravitational
redshifts.  For stars where independent information is available for
either the mass or the radius, this application of general relativity
directly determines the other quantity.  Applicable examples are: 1)
measuring the radii of solar metallicity main sequence stars for which the
mass - luminosity relation is well known, 2) measuring the radii for stars
where model atmospheres can be used to determine the surface gravity (the
mass to radius squared ratio), 3) refining the mass - radius relation for
main sequence stars, and 4) measuring the change in radius as stars evolve
off the main sequence and up the giant branch.

\end{abstract}


\section{Introduction}

Einstein (1911) predicted a gravitational redshift for light escaping the
Sun equivalent to 0.636 km s$^{-1}$ based on his General Theory of
Relativity (Einstein 1916).  Early pioneering work (Evershed 1931)
demonstrated that absorption lines near the solar limb were systematically
redshifted, although the measurements did not entirely agree with
Einstein's theory as the experiment was too difficult for the
instrumentation of the time.  Subsequent work (see Vessot et al. 1980;
LoPresto et al. 1991; Krisher, Morabito, \& Anderson 1993) yielded results
which, within the errors (now 1\%), were indistinguishable from Einstein's
predictions.  While measuring the solar gravitational redshift has been a
better means of testing General Relativity than of measuring the mass of
the Sun, subsequent gravitational redshift work has been aimed primarily
at measuring the masses of white dwarf stars in binaries (e.g. Adams 1925;
Wegner 1980; Wegner \& Reid 1991) and open clusters (for the early work see
Greenstein \& Trimble 1967 and Trimble \& Greenstein 1972; for a modern
work see Reid 1995).  The high surface gravities of white dwarfs produce
redshifts of $\approx$ 30 km s$^{-1}$ (Wegner \& Reid 1991).  The
procedure is to measure the radial velocity of a white dwarf relative to a
known systemic velocity, such as that of an open cluster or binary
system.

Modern instrumentation (see Walker 1992) now makes it possible to extend
gravitational redshift measurements to a new regime and measure the
gravitational redshifts of main sequence stars.  Such measurements would
directly determine the mass - radius ratio ($M/R$) for these stars.  While
the mass of main sequence stars as a function of luminosity, color or
spectral type is well-determined for solar metallicity stars from binary
star studies (Popper 1980), it is not well-determined for metal-poor or
metal-rich stars.  Furthermore, stellar radii have not been measured for
most types of stars.  Generally, the techniques used to determine stellar
radii measure the angular size of a star (e.g. lunar occultation and
interferometric techniques, see McAlister (1985)), and thus also require a
distance determination.  These techniques also generally favor nearby
giant stars.  Detailed studies of double-line spectroscopic eclipsing
binaries (e.g. Nordstr\"om \& Johansen 1994) can yield stellar radii
without the need for distance measurements, but few such eclipsing
binaries exist.  Besides being a fundamental parameter, the radii of stars
with convective outer layers cannot be determined adequately from
currently understood physics (see below).

Following, I suggest two methods for measuring $M/R$ for main sequence
stars in open clusters.  Both methods are based on the very low velocity
dispersions in many open clusters, for instance $\sigma$ = 0.44 $\pm$ 0.04
km s$^{-1}$ in the Hyades (Zhao \& Chen 1994) and $\sigma$ = 0.48 $\pm$
0.09 km s$^{-1}$ in M67 (Mathieu 1985).

\section{Method 1: Relative measurement along the Main Sequence}

This method uses the intrinsic, but slow, variation of the mass - radius
relation along the main sequence to measure the relative increase in $M/R$
as a function of increasing stellar mass.  The gravitational redshift is
given by Greenstein \& Trimble (1967) as:

\begin{equation}
K = 0.635 (M/R)
\end{equation}

\noindent
where K = gravitational redshift in km s$^{-1}$, and $M$ and $R$ are in
solar units.  (Note that the coefficient here is slightly different from
the 0.636 km s$^{-1}$ value predicted by Einstein.  This is simply a
commonly-used round-off of the Einstein value.)

The mass - radius relation (Mihalas \& Binney 1981) can be approximated
as:

\begin{equation}
R \sim M^{0.7}
\end{equation}

\noindent
again with $M$ and $R$ in solar units, yielding:

\begin{equation}
K = 0.635 M^{0.3}
\end{equation}

\noindent
Thus a G2 V star (1.0 $M\sun$) should have K = 0.635 km s$^{-1}$ (and the
Sun does, see above) and a B3 V star ($\approx$ 10 $M\sun$) should have K
= 1.267 km s$^{-1}$.  In practice one would determine the radial
velocities of all available cluster members, remove binary stars, and
correlate the measured stellar velocities with color, effective
temperature, or luminosity, which themselves correlate with mass.  In the
example cited here, one is attempting to measure a difference of $\approx$
0.6 km s$^{-1}$ (G2 V to B3 V) within a system with a velocity dispersion
of $\approx$ 0.5 km s$^{-1}$.  Modern velocity measurements can be made
with a precision significantly better than 0.5 km s$^{-1}$ for F-type and
later stars (Duquennoy, Mayor, \& Halbwachs 1991), while for A-type and
earlier stars the paucity of absorption lines and the generally high
rotation velocities make it difficult to achieve a precision better than
1.0 km s$^{-1}$ (Morse, Mathieu, \& Levine 1991).  Thus for most cluster
stars the cluster dispersion dominates the random errors.  Systematic
errors which are a function of stellar surface temperature present a
larger observational challenge, however.  Such systematic errors arise
primarily from line blends which change in relative strength as a function
of atmospheric temperature, inducing small, but real, line centroid
shifts.  There appears to be, for example, a systematic dependence on
temperature in the radial velocities measured with the CfA speedometers
(Latham 1995) at the few tenths of a km s$^{-1}$ level.  This systemic
error is most likely due to the small wavelength range used, so that
mismatches in blends do not average out adequately.  Increased wavelength
coverage should greatly reduce this source of systematic error.  Increased
resolution may also help to de-blend these lines.

I assume that the errors in color, effective temperature, or luminosity
can be made small enough so that the measurements are limited by the
velocity measurements.  This is reasonable since internal photometric or
spectroscopic temperature errors can be reduced to $\leq$ 1\% per star,
and with $\leq$ 2\% external accuracy (Young 1993).  A suitably chosen
color index, such as V-I, will span a magnitude between a B3 V and a G2 V
star, for example, and the effective temperature difference is a factor of
three.

Assume for this discussion that insignificant systematic errors can be
achieved\footnote{If systematic errors cannot be greatly reduced by
increased wavelength coverage, then this problem can be inverted and the
gravitational redshift effects discussed here should be {\it assumed} to
determine the level of systematic error still resident in the data.}. Then
high-quality velocity measurements will yield errors $\approx$ 0.3 km
s$^{-1}$ for F-type and later stars and $\approx$ 1.0 km s$^{-1}$ for
A-type and earlier stars.  Combining these with the $\approx$ 0.5 km
s$^{-1}$ velocity dispersion yields $\sigma_{late-type}$ $\approx$ 0.6 km
s$^{-1}$ and $\sigma_{early-type}$ $\approx$ 1.1 km s$^{-1}$.  The
uncertainty for the late-type stars is the same size as the gravitational
redshift difference across this mass range, $\approx$ 0.6 km s$^{-1}$.  In
this example the goal is the slope in a velocity - mass (or color, etc.)
diagram.  The early-type and late-type stars each cover approximately a
factor of 3.3 in mass range in this example, and so an approximation of
the number of stars needed for a reliable measurement comes from acquiring
a ratio (=3.4) of early-type to late-type stars to give each mass bin the
same total error, then increasing the number of stars to the point where
$\surd N$ statistics yield a meaningful result.  Thus $\approx$ 12 stars
along the main sequence mass range from G2 V to A0 V and 40 from A0 V to
B3 V would produce a 5 $\sigma$ (=0.12 km s$^{-1}$) measurement,
statistically determining the ratio of $M/R$ at 10 $M\sun$ to $M/R$ at 1
$M\sun$ to within 20\%.

This method would yield $M/R$ for main sequence stars spanning a range
in mass.  If the cluster distance, reddening and metallicity are well
known, and if the metallicity is near solar, then the mass - luminosity
relation can be used to give stellar masses, from which stellar radii can
be determined.  Alternatively, model atmosphere fits to the spectra of
individual stars can yield the surface gravity, which is proportional to
$M/R^2$, and again radii can be determined\footnote{See, however,
Bergeron, Liebert \& Fulbright 1995 for a discussion of the small, but
significant, inconsistencies between gravitational redshift and
atmospheric mass determinations.}.

As a final note to this method, the mass - radius relation for main
sequence stars is not entirely a power law, but has some subtle structure,
especially for the lowest mass stars with spectral types later than M3.
Careful work along a number of cluster main sequences to very faint
magnitudes may eventually be possible, allowing a refined mass - radius
relation.

\section{Method 2: Relative measurement from the main sequence turn-off
through the giant branch stars}

This method uses the rapid increase in radius at a nearly constant mass as
stars evolve off the main sequence and up the giant branch.  Typically
stellar radii double from the main sequence turn-off to the blue hook,
then remain nearly constant across the subgiant branch to the base of the
red giant branch (see Bressan et al. 1993).  Once on the giant branch,
stellar radii rapidly increase.  Since stellar surface temperatures differ
very little as stars increase their radii and move up the giant branch,
this technique is not as sensitive to the systematic velocity errors as
Method 1.  This method employs velocities in luminosity intervals to
provide stellar radii at those points in the HR diagram.

{}From the arguments presented above, it is easy to see that a cluster with
a turn-off at $\approx$ 1.6 $M\sun$ (i.e. at F0 and with age $\approx$ 2
Gyrs) would exhibit a gravitational redshift velocity falling from
$\approx$ 0.73 km s$^{-1}$ on the main sequence to essentially 0 km
s$^{-1}$ somewhere along the giant branch.  With $\sigma_{late-type}$
$\approx$ 0.6 km and $\approx$ 18 stars per luminosity interval (i.e. in
each of the turn-off region, the subgiant branch, and the various
luminosity bins up the giant branch), for example, velocities would be
determined to $\approx$ 0.15 km s$^{-1}$, providing a 5 $\sigma$
measurement across the applicable luminosity range, and measuring stellar
radii to within 20\% on the main sequence and to within 40\% at the point
where the radius has doubled from its main sequence value.  A
determination of the run of radii as a star evolves would provide a
valuable constraint on stellar models, since current theory cannot
precisely determine stellar radii for most stars.  This weakness arises
because the physics of convection is poorly understood, and thus one
cannot calculate from first principals the radius of any star with surface
convection.  The convection theory used in most stellar models, known as
mixing length theory, is a simple parameterization of convective cells
rising some ratio of the pressure scale height.  The single parameter of
this model is fit to the standard solar model, but detailed studies (e.g.
Taylor 1986) of its applicability to other stars provide significant
evidence that this theory is too simplistic.

This technique potentially could be inverted if the radii of some subset
of the cluster stars were known, e.g. by lunar occultation measurements,
in which case the mass of the turn-off stars, and thus their age, could be
measured.  This would be easiest for very young clusters with higher mass
turn-offs, since stellar lifetimes are a steep function of mass, though
the number of high mass stars available would be small.

\acknowledgments
I am grateful to the referee, Timothy Krisher, for comments and
suggestions which helped me improve and clarify this paper.  I would like
to thank Sam Barden, Stephane Charlot, Pierre Demarque, Dave Latham, Bob
Mathieu, Caty Pilachowski, and Ata Sarajedini for helpful discussions.
This work was partially supported by a grant from the Edgar P. and Nona
B.  McKinney Charitable Trust.

\end{document}